\documentclass{article}

\usepackage[english]{babel}
\usepackage{multirow}
\usepackage{diagbox}
\usepackage[letterpaper,top=2cm,bottom=2cm,left=3cm,right=3cm,marginparwidth=1.75cm]{geometry}
\usepackage{float}
\usepackage{amsmath}
\usepackage{graphicx}
\usepackage[colorlinks=true, allcolors=blue]{hyperref}

\title{A novel methodological framework for analyzing the momentum effect in tennis singles }
\author{Chang Du$^1$, Caiya Zhang$^2$, Likai Zhou$^1$\\
\it {$^1$ School of Data Sciences, Zhejiang University of Finance $\&$ Economics}\\
\it {$^2$ Department of Statistics and Data Science, Hangzhou City University}
\thanks{Corresponding author: Caiya Zhang, email address: zhangcy@hzcu.edu.cn}}

\date{}
\begin{document}
\maketitle

\begin{abstract}

This paper proposes a novel methodological framework for analyzing momentum effects in tennis singles. To statistically substantiate the existence of momentum, we employ Chi-squared independence tests for the contingency table. Assuming momentum is present, we develop a momentum metric based on the entropy weight method. Subsequently, we apply a CUSUM control chart to detect change points within the derived momentum series and define a relative distance measure to quantify the intensity of momentum shifts. Furthermore, we construct a predictive model utilizing a Back Propagation neural network (BP) optimized by a Particle Swarm Optimization (PSO) algorithm. The importance of predictive features is analyzed via SHAP values. An empirical analysis applying this framework to data from the 2023 Wimbledon Men's Singles yields the following key findings: (1) Statistical evidence significantly supports the existence of momentum in tennis singles. (2) Incorporating momentum characteristics substantially enhances point outcome prediction performance. (3) The BP+PSO model demonstrates competitive advantages over alternative machine learning algorithms, including Random Forest, Support Vector Machines, and logistic regression. (4) SHAP value analysis identifies an athlete's unforced errors, winning shots, the momentum metric, and the momentum shift intensity as the four most critical features for predicting point outcomes.
\end{abstract}

\section{Introduction}
In competitive sports, athletes who have consecutive wins often maintain peak performance in subsequent matches, while those who endure consecutive losses tend to struggle further. This well-documented phenomenon is widely known as the momentum effect or the "hot hand" effect. Despite its widespread intuitive acceptance, the existence and underlying mechanisms of this phenomenon have long been a subject of intense debate within academia. A deeper understanding of the momentum effect can assist athletes in better adjusting their mentality, allocating physical energy, and devising strategies during training and competitions. Additionally, it can provide coaches, media, and fans with more reliable match analyses, aiding in better decision-making and a more informed viewing experience.

Early studies on the momentum effect mainly focused on basketball. In 1985, Gilovich et al. analyzed shooting data from the Boston Celtics and Cornell's varsity team using conditional probability. Their results showed that while players' beliefs about streakiness influenced their predictions, there was no statistically significant evidence of actual performance changes,leading them to reject the "hot hand" hypothesis. Subsequent research, such as that conducted by Adams (1992) and Vergin (2000), further solidified the aforementioned conclusion. However, their static research methodologies were inadequate in capturing the dynamic and ephemeral characteristics of momentum. In stark contrast to these early findings, an increasing number of studies have provided empirical evidence to support the existence of momentum effects in basketball. Mace et al.(1992) demonstrated that strategic timeouts disrupted opponents’ psychological momentum in NCAA tournaments. Bocskocsky et al. (2014) quantified momentum impact on NBA shooting percentages through large-scale spatial analytics. Miller and Sanjurjo (2018) specifically re-analyzed the data from Gilovich and Vallone (1985). They pointed out some flaws in the statistical methodology used by Gilovich and Vallone (1985) and revealed a subtle but significant bias in common metrics measuring how current outcomes depend on prior sequential occurrences in sequential data. After correcting for the bias, it was found that the long-held conclusions of the classic studies were reversed, indicating that there is, in fact, evidence supporting the existence of the "hot hand " effect.
Latins and Nessen (2021) investigated the hot hand using detailed data on free throws and field goals attempts for 12 NBA seasons. They found a small hot hand for free throws, which more than doubles for longer streaks of made free throws. Both offenses and defenses respond to field goals, but a made field goal does not change the probability that a player makes his next field goal attempt, and longer streaks of made field goals reduce the probability that a player makes his next field goal attempt.

With the widespread application of data analysis techniques in competitive sports, research on the momentum effect has been expanded to cover other sporting competitions like horseshoes (Smith, 2003), golf (Livingston, 2008), baseball (Young et al., 2020), and football (Gauriot and Page, 2023). Tennis, as a sport that combines competitiveness, individuality, and psychological confrontation, presents an ideal setting for studying the momentum effect. Especially in Grand Slam tournaments,  players have to compete in multiple rounds under high-pressure conditions, making their competitive states highly volatile. In 1997, Jackson and Mosurski, by analyzing tennis match data, they explored the association between psychological momentum and match outcomes. The results indicated that momentum fluctuations during matches are not random but are influenced by specific factors, and there existed a certain correlation between momentum changes and players' success. Klaassen and Magnus (2003) verified that in Wimbledon tennis matches winning the previous set had a positive impact on the probability of winning in the subsequent set. Malueg and Yates (2010), based on ATP (Association of Tennis Professionals) data from 2004-2013, employed a Markov chain model and dynamic probability estimation to find that the outcome of the first set had a significant influence on the subsequent performance of players. Moss and O’Donoghue (2015), using the data from men's singles matches at the US Open from 2002-2013, conducted conditional probability analysis and established a logistic model, discovering that winning or losing streaks significantly influenced the winning probabilities in subsequent matches. Building upon Moss and O'Donoghue's (2015) dataset, Depken et al. (2022, 2023) identified two distinct forms of momentum: tactical momentum and psychological momentum, both present in best-of-three and best-of-five tennis singles matches. Their analysis further revealed the dynamic changes and reversals of these two types of momentum during the course of matches. 

It is noteworthy that among the aforementioned studies, the primary focus has been on confirming the presence of momentum in tennis competitions. However, a systematic framework for examining the influence mechanisms of the momentum effect has yet to be established. For instance, there has been a notable absence of in-depth research concerning how to quantify the extent of momentum and how to apply it to improve the prediction of outcomes. The principal objective of our paper is to put forward a well-structured framework for analyzing the momentum effect in tennis singles. To examine the presence of the momentum effect, we construct a contingency table tallying the frequency of winning or losing the subsequent point after winning streaks. We then conduct a Chi-squared independence test and conditional probability analysis  to statistically verify the existence of momentum. 
 Assuming momentum is present, we subsequently address the quantification challenge. Prior research has predominantly treated momentum as a qualitative construct or resorted to simplified correlation models, which frequently fell short in capturing the multifaceted and dynamic essence of momentum within competitive sports. In the present study, we employ the entropy weight method to construct a comprehensive variable that encompasses key features pertinent to players' performance. This newly developed indicator not only captures the direct influence of momentum on match results but also accounts for its broader dynamic characteristics. Experience shows that momentum often s between athletes during a competition. We use the cumulative sum (CUSUM) control chart to find the change points of momentum. A relative distance is also defined to indicate the intensity and direction of momentum transfer, which helps us gain a deeper understanding of momentum dynamics. In the subsequent step, we harness the capabilities of a BP neural network to develop a prediction model, with indicators including momentum measurement, change-point detection results of momentum, and momentum - shifting intensity in the input layer. To optimize the model, we adopt the Particle Swarm
Optimization (PSO) algorithm, renowned for its global search capability. The parallel search mechanism of this algorithm not only accelerates convergence but also enhances the optimization efficiency and robustness of the model. The empirical study results demonstrate that the proposed momentum measurement
 and the intensity of momentum transfer   hold substantial value in predicting the outcome of a point. Additionally, compared to some traditional machine learning algorithms such as Random Forest, SVM and Logistic Regression, the BP+PSO model exhibits distinct competitive advantages.
 
The rest of the paper is organized as follows. Section 2 proposes a novel methodological framework for momentum effect analysis. Section 3 conducts an empirical study using the data from 2023 Wimbledon Men’s Singles Championships. Some conclusions and discussion are given in Section 4.

\section{Methodological framework}

In tennis singles, men's matches follow a best-of-five sets format, while women's matches use a best-of-three sets structure. Each set consists of multiple games, and each game is made up of individual points. To dissect the momentum effect, we should exam athletes' micro-level performance across individual points. The two players are designated as Player 1 and Player 2. In subsequent sections, all analytical results and mathematical formulations principally concern Player 1, except where explicitly stated otherwise.

\subsection{Tests for the existence of momentum}

Define events
 \(W\) as winning a single point, \(L\) as losing a single point,  \(W_i\) as a winning streak of  \(i\) consecutive points, 
\(L_i\) as a losing streak of 
\(i\) consecutive points .

Let
\(n_{i.}\) be the total number of \(W_i\) streaks, \(n_{i1}\) be the subcount of steaks extended to \(W_{i+1}\) (next point \(W\)), \(n_{i2}\) be the subcount of streaks terminated at \(L\) or truncation (next point \(L\) or the game is over), such that \(n_{i.}=n_{i1}+n_{i2}\). 

Given \(W_i\), we get the transition probabilities   
\[
P(W_i\rightarrow W) =P(\text{Extension}|W_i)= \frac{n_{i1}}{n_{i.}},
\]
\[
P(W_i\rightarrow L ~\text{or over}) = P(\text{Termination}|W_i)=\frac{n_{i2}}{n_{i.}}.
\]

For illustration, consider a sample sequence including 14 points: 
\[W - W - L - W - W - W - L - W - L - W - W - L - W - W.\]
\begin{itemize}
   \item For $i=1$ ($P(\text{Extension}|W_1)$):
 There are 5 streaks of $W_1$ at points $t=1,4,8,10,13$, respectively,  which means $n_{1.} = 5 $. Among these, 4 streaks are extended to 2 or 3 consecutive wins, 1 streak is terminated at $L$, so $n_{11} = 4$.The estimated probability $P(\text{Extension}|W_1) = 4/5 = 0.8$.

    \item For $i=2$ ($P(\text{Extension}|W_2)$): There are 4 streaks of $W_2$ at points $t=1-2, 4-5, 10-11, 13-14$, respectively, which means $n_{2.} = 4 $. Among these, there are 3 streaks of $W_2$ are terminated, where 2 of them are followed by $L$ and 1 of them is game over. So the estimated probability $P(\text{Extension}|W_2) = 1/4 = 0.25$.
    
    \item For $k=3$ ( $P(\text{Extension}|W_3)$): There is only 1 streak of $W_3$ at points $t=4-6$. Since the next point followed this streak is  $L$, the estimated probability $P(\text{Extension}|W_3) = 0$.
\end{itemize}
For detailed information, see Table \ref{Tab2-0}.
\begin{table}[H]
    \centering
    \caption{\label{Tab2-0}\centering{Winning Streak Transition Frequency Table }}
    \begin{tabular}{c|c|c|c|c}
    \hline
    \textbf{Streak Length } ($i$) & \textbf{Streak Positions}($t$) & \textbf{Next Point} & $n_{i.}$ & $n_{i1}$\\
    \hline
    $i=1$ & $t=1,4,8,10,13$ & $W,W,L,W,W$ & 5&4(t=1,4,10,13) \\
    $i=2$ & $t=1-2,4-5,10-11,13-14$ & $L,W,L,\text{ truncation}$ & 4&1(t=4-5)\\
   $i=3$ & $t=4-6$ & $L$ & 1&0\\
          \hline
    \end{tabular}
\end{table}

If momentum is absent in tennis singles competitions, the sequence of points follows a first-order Markov chain. Under this hypothesis, the length of a winning streak and the outcome of next point are statistically independent. This implies the conditional probabilities
$\{P(\text{Extension}|W_i), 1\le i \le k\}$ 
should be constant across all streak lengths $i$, where $k$ denotes the maximum observed streak length. To statistically validate momentum existence, we implement a contingency table independence test for streak length versus point outcome. Consider the null hypothesis
\[
H_0: P(\text{Extension}|W_1) = P(\text{Extension}|W_2) =  \cdots = P(\text{Extension}|W_k),
\]
and the alternative hypothesis 
\[
H_a: P(\text{Extension}|W_1),~ P(\text{Extension}|W_2),~  \cdots, P(\text{Extension}|W_k) ~~\text{are not all equal.}
\]
Based on the point outcomes, we build a contingency table similar to Table \ref{Tab2-1}.

\begin{table}[H]
    \centering
    \caption{\label{Tab2-1}\centering{Contingency table for the point outcomes  }}
    \begin{tabular}{|c|c|c|c|}
    \hline
    \diagbox{\textbf{Streak}}{\textbf{Subsequent point}} & \textbf{Extension} & \textbf{Termination} & $n_{i.}$ \\
    \hline
    $W_1$ & $n_{11}$ & $n_{12}$ & $n_{1.}$ \\
    $W_2$ & $n_{21}$ & $n_{22}$ & $n_{2.}$ \\
   
   \vdots &\vdots & \vdots & \vdots \\
    $W_k$ & $n_{k1}$ & $n_{k2}$ & $n_{k.}$ \\
    \hline
    $n_{.j}$ &$n_{.1}$ & $n_{.2}$ & $n$ \\
    \hline
    \end{tabular}
\end{table}
In Table \ref{Tab2-1}, \( n_{ij} \)  stands for the observed frequency. The expected frequency is given by  \( \hat{n}_{ij}=\frac{n_{i.}n_{.j}}{n} \)
 , where $n_{i.}=n_{i1}+n_{i2}, n_{.j}=\sum_{i=1}^k n_{ij}$ and \( n=\sum_{i=1}^k\sum_{j=1}^2n_{ij} \).
When all expected frequencies satisfy $\hat{n}_{ij} \geq 5$ and the total number of winning streaks $n \geq 50$, we apply Pearson's Chi-squared test. 
The test statistic is
\[
\chi^2=\sum_{i=1}^k\sum_{j=1}^2{\frac{(n_{ij}-\hat{n}_{ij})^2}{\hat{n}_{ij}}}.
\]
 Under the null hypothesis, $\chi^2$ asymptotically follows a $\chi^2$-distribution with $k-1$ degrees of freedom. Let $\chi_0^2$ denote the observed test statistic. If the $p$-value $P(\chi^2 > \chi_0^2)$ falls below the significance level $\alpha$, we reject the null hypothesis, providing statistically significant evidence for momentum existence.

When either $\hat{n}_{ij} < 5$ for any cell or $n < 50$, we recommend the Fisher-Freeman-Halton exact test, with theoretical foundations detailed in Freeman \& Halton (1951).

\subsection{Measurement of momentum based on Entropy Weight Method}
Given the verified existence of momentum, the subsequent question arises as to how to quantify it. Assume that there exist $m$ important features $x_1, x_2,\cdots, x_m$, which are relevant to the players' performance at each point, encompassing technical skills, stability and error control, physical effort and shot intensity etc.
We employ the Entropy Weight Method (EWM) to construct a composite variable as a measure of momentum.
EWM is a well-established approach grounded in information entropy principles for multi-index comprehensive evaluation. The basic idea of EWM is to allocate weights based on the entropy values of each indicator. Specifically, the greater the entropy value of an indicator, the more informative it is, consequently, the larger the weight it should be assigned, and vice versa. 

For a fixed match comprising $T$ points, let $x_{i}^{(t)}$ denote the value of the $i$-th feature for Player 1 at point $t$, where $t \in \{1, 2, \ldots, T\}$ and $i \in \{1, 2, \ldots, m\}$.
 The procedures for calculating the weights using EWM are outlined as follows.

\textbf{Step 1 Feature Standardization}

\begin{itemize}
    \item \textbf{Positive feature} For features where larger values are preferable, such as winning percentage and points scored,
    \[
    z_{it} = \frac{x_{it} - x_{i,\min}}{x_{i,\max} - x_{i,\min}},
    \]
    
    \item \textbf{Negative feature} For features where smaller values are preferable, such as error rate and unforced errors,
    \[
    z_{it} = \frac{x_{i,\max} - x_{it}}{x_{i,\max} - x_{i,\min}},
    \]
\end{itemize}

where $x_{i,\max} = \max_t \{ x_{it} \}$ and $x_{i,\min} = \min_t \{ x_{it} \}$.

\textbf{Step 2 Feature Normalization}

To obtain the proportion of each feature, define
\[
p_{it} = \frac{z_{it}}{\sum_{t=1}^T z_{it}}.
\]

\textbf{Step 3 Calculate the Entropy Value}

The entropy value of each feature is 
\[
e_i = -\frac{1}{\ln(T)} \sum_{t=1}^T p_{it} \ln(p_{it} + \epsilon),
\]
where $\epsilon$ is a small positive constant to prevent the occurrence of $\ln(0)$.

\textbf{Step 4 Calculate the Weight}

The weight of each feature is given by:
\[
w_i = \frac{1 - e_i}{\sum_{i=1}^m (1 - e_i)}.
\]

Then the measurement of the momentum  for Player 1 at point $t$  is defined by the comprehensive variable
\[
 M_t = \sum_{i=1}^m w_i z_{it}, t=1,\cdots, T.
\]
\subsection{CUSUM-based change point detection of momentum series}
Empirical studies have confirmed that the competitive momentum undergoes dynamic shifts and abrupt reversals between opponents during  match play(see Depken et al. 2022, 2023). Hence, it is necessary to identify the change points of momentum series. 
 CUSUM control chart proposed by Page (1954) is a statistical tool widely used for change point detection. 
 
The basic idea of CUSUM method is to monitor the cumulative deviation of each data point from a reference mean. By accumulating these deviations over time, the CUSUM method can detect even subtle changes in the mean performance with high sensitivity. Assume that, in a fixed match, \( \mu \) represents the reference mean value of $M_t$. we calculate the cumulative sum of deviations from \( \mu \) for each point. Once this cumulative sum exceeds a predefined threshold, it indicates a significant shift in the underlying performance trend.

Let \( C_t \) represent the cumulative sum of deviations at point $t$, defined as follows.

\[
C_t = M_t - \mu + C_{t-1} - d,
\]

where
\begin{itemize}
    \item \( C_0 = 0 \), which initializes the cumulative sum to zero.
    
    \item \( d \) is a drift parameter used to control the accumulation rate of deviations.     
\end{itemize}
When the absolute value of $C_t$  exceeds the predefined positive threshold \( h \), point \( t \) is identified as a change point. Specifically, a change point where $C_t>h$ signifies an increase in the player's winning probability. We refer to such a point as a \textit{positive change point}. Conversely, a change point where $C_t<-h$ indicates a decrease in the player's winning probability. We refer to this as a \textit{negative change point}. Define the sign feature \(CP_t\) as follows. 
\[
CP_t=\begin{cases}
+1, & \text{for}~C_t>h,\\
0, & \text{for}~-h\leq C_t\leq h,\\
-1, & \text{for}~C_t<-h.
\end{cases}
\]

The drift parameter  \( d \) and  threshold \( h \) play a crucial role in determining the performance of the CUSUM algorithm. A smaller drift value  enhances the algorithm's sensitivity to minor changes, while a larger value makes it more robust against transient fluctuations. Conversely, a lower threshold increases the number of detected change points, improving the detection of short-term variations, whereas a higher threshold reduces false alarms but may miss subtle shifts. In practice, these parameters should be carefully tuned based on the data characteristics. Iterative testing, combined with visual analysis, can help optimize the trade-off between sensitivity and robustness, ultimately improving the accuracy and efficiency of change point detection.

\subsection{ Dynamic assessment of momentum shift intensity }
Consider a momentum time series $\{M_t\}_{t=1}^T$ over a match including $T$ points, with $N$ detected change points at $\{t_1, t_2, \ldots, t_N\}$. 
Denote \(\{D_i, 1\le i \le N\}\) as the interval durations between two adjacent change points, that is,
\[D_1=t_1, D_i=t_i-t_{i-1}, i=2,3,\cdots, N.\]
To quantify dynamic momentum transitions and evaluate the intensity of shift events during a match, we propose a {\it relative distance} metric, denoted by $V_t$, which captures both the direction and the magnitude of momentum shifts. 
\begin{itemize}
    \item \textbf{At change points $\{t_1, t_2, \ldots, t_N\}$.}
The sign of $V_{t_i}$ is determined by $CP_{t_i}$ (defined in Section 2.3). A positive $V_{t_i}$ indicates a positive momentum shift, signifying Player A's dominance (e.g., winning streaks or tactical adjustments), while a negative $V_{t_i}$ reflects Player B's counterattack momentum (e.g., break point conversions or disruptive strategies). The magnitude of $V_{t_i}$ scales inversely with interval duration $D_i$: shorter intervals correspond to higher intensity shifts. We therefore define:
    \[
    V_{t_i} = CP_{t_i} \times \left(\frac{D_{max}}{D_i}\right),
    \]
    where $D_{max} = \max(D_1, D_2, \ldots, D_N)$. 

    \item \textbf{Between change points $t \in (t_{i-1}, t_i)$.}
     $V_t$ is computed through linear interpolation between $V_{t_{i-1}}$ and $V_{t_{i}}$,
    \[
    V_t = \frac{V_{t_i} - V_{t_{i-1}}}{t_i - t_{i-1}} \times (t - t_{i-1}) + V_{t_{i-1}}.
    \]
    This ensures both smooth momentum intensity transitions and directional continuity across intervals.

    \item \textbf{Boundary Conditions.} $V_t$ is also computed through linear interpolation. 
     \begin{itemize}
     \item Initial phase ($t < t_1$):
        \[
        V_t = \frac{V_{t_1}}{t_1} \times t.
        \]
        \item Terminal phase ($t > t_N$):
       \[
        V_t = V_{t_N} - \frac{V_{t_N}}{T - t_N} \times (t - t_N).
        \]
\end{itemize}
    The above two formulations maintain the directional consistency in momentum accumulation before the first change point or decay processes after the last change point.
    \end{itemize}

\subsection{Prediction based on BP neural network with PSO algorithm}

 To develop an accurate and robust predictive framework, we implement a hybrid computational intelligence approach that synergistically combines a Back Propagation (BP) neural network  with Particle Swarm Optimization (PSO). This integrated methodology offers distinct advantages for modeling complex nonlinear relationships in competitive momentum analysis.
 
\begin{figure}[h]
\centering
\includegraphics[width=0.6\linewidth]{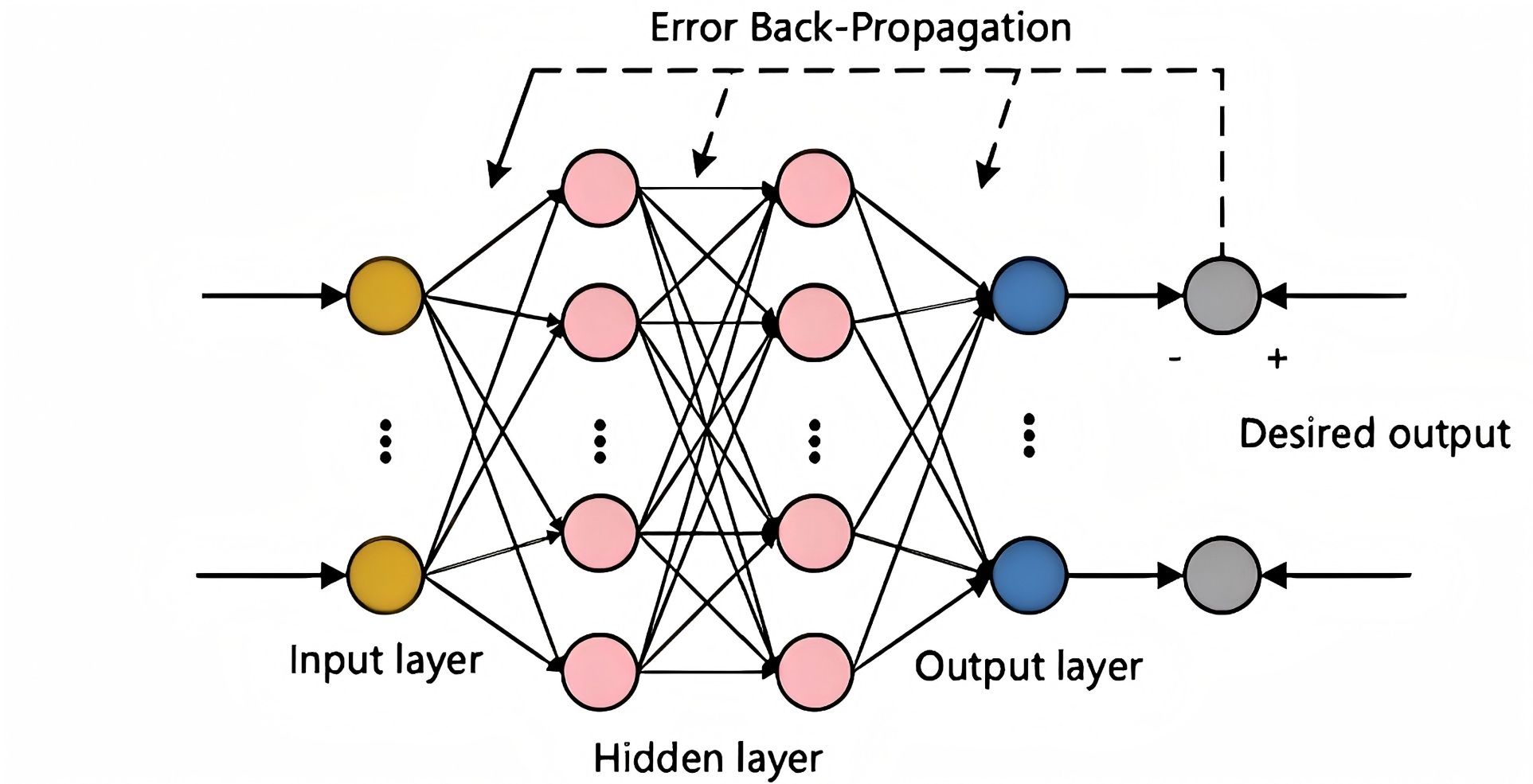}
\caption{\label{fig:BP} BP neural network flow chart }
\end{figure}

The BP neural network, introduced by Rumelhart, McClelland, and colleagues in 1986, is a supervised learning algorithm celebrated  for its  exceptional nonlinear mapping capabilities.  By iteratively adjusting the weights of inter-neuronal connection to learn input-output relationships, it boasts the universal approximation property enabling it to model any continuous function. Additionally, it exhibits remarkable adaptability to high-dimensional data.
The BP neural network has been widely applied across various domains, including image recognition, natural language processing, and predictive analytics. Figure \ref{fig:BP} illustrates the flow chart of the BP neural network.  

However, the Backpropagation (BP) neural network is prone to becoming trapped in local minima, particularly in high-dimensional feature spaces where conventional gradient descent methods often face difficulties in converging to the global optimum. This problem stems from the fact that the loss landscape in these spaces is typically highly non-convex, featuring a multitude of local minima and saddle points that can easily derail the optimization process.

To address this challenge, we adopt the Particle Swarm Optimization (PSO) algorithm, which was introduced by Eberhart and Kennedy in 1995, to optimize the BP neural network.

 PSO is a global optimization technique inspired by the collective behavior observed in bird flocks or fish schools. In these natural systems, individuals within the group exchange information to collectively  identify the optimal solution. In the PSO algorithm, each particle within the swarm represents a potential solution in the search space. The entire swarm cooperatively searches for the optimal solution by iteratively updating the position and velocity of each particle. During each iteration, a particle's position is adjusted based on its personal best solution 
 (\(p_{best}\)) and the global best solution 
 (\(g_{best}\)) identified by the entire swarm up to that point.
\[
v_{i}(k+1) = \omega v_{i}(k) + c_1r_1(p_{best} - p_{i}(k)) + c_2r_2(g_{best} - p_{i}(k))
\]
where
\begin{itemize}
    \item \(v_{i}(k)\) is the velocity of particle \(i\) at iteration \(k\);
     \item \(p_{i}(k)\) is the particle position at  iteration \(k\);
    \item \(\omega\) is the inertia weight, which controls the influence of the previous velocity on the current movement;
    \item \(c_1, c_2\) are learning factors, determining the influence of personal and global best positions;
    \item \(r_1, r_2\) are random values between 0 and 1;
    \item \(p_{best}\) is the personal best position of particle \(i\);
    \item \(g_{best}\) is the global best position found by the swarm.       
\end{itemize}
The position \(p_i(k+1)\) of the particle at iteration 
\(k+1\) is updated with the new velocity \(v_i(k+1)\). This iterative process persists until the optimization criteria are satisfied. 

To quantitatively analyze the impact of the momentum effect on prediction outcomes, four distinct input layer scenarios are integrated into the BP neural network model. 
(1){\bf{Base}}: Represents the basic feature set. (2){\bf{Base+M}}: Incorporates the momentum features $M$ into the base feature set. (3){\bf{Base+M+CP}}: Adds the change point label feature $CP$ to the previous configuration. (4){\bf{Base+M+CP+V}}: Further includes the intensity of momentum shift $V$. A comparative evaluation of prediction results is performed using four classification metrics: precision, recall, F1-score, and the Area Under the Receiver Operating Characteristic  Curve (AUC). 

To evaluate the contribution of each feature to the predictive performance, we calculate SHAP values (Lundberg \& Lee, 2017) based on Shapley's cooperative game theory (Shapley, 1953) to quantify feature importance. 
\[
\phi_i = \sum_{S \subseteq F \setminus \{i\}} \frac{|S|!(|F|-|S|-1)!}{|F|!} [f(S \cup \{i\}) - f(S)]
\]
where
\begin{itemize}
\item \(\phi_i\): The SHAP value of feature \(i\), representing its contribution to the model's prediction.
\item \(F\): The set of all features.
\item \(S\): A subset of features that does not include feature \(i\), i.e., \(S\) is a subset of \(F\) with feature \(i\) removed.
\item \(f(S)\): The model's output when using the feature subset \(S\) for prediction.
\item \(f(S\cup{i})\): The model's prediction when feature \(i\) is added to the feature subset S.
\item \(|S|\): The size of the feature subset \(S\), i.e., the number of features in subset \(S\).
\item \(|F|\): The size of the full feature set \(F\), i.e., the total number of features.
\end{itemize}

 \section{Empirical analysis}

The dataset employed in this study is sourced from the 2024 Mathematical Contest in Modeling (MCM) (Problem C: Momentum in Tennis). It comprises comprehensive point-by-point records from the 2023 Wimbledon Championships Men's Singles tournament, covering 31 matches after the first two rounds. This dataset includes the final between Carlos Alcaraz and Novak Djokovic (identified as match$_{-}$id: 2023-wimbledon-1701). The complete data package with variable definitions is publicly available via the competition portal(
https://www.mathmodels.org/Problems/2024/MCM-C/index.html). According to the tournament regulations, to win a set, a player must win at least 6 games and be ahead by at least 2 games. For the first four sets, the tie-breaker is won by the first player to reach 7 points with at least a 2-point lead. In contrast, the fifth set tie-breaker requires a player to reach 10 points with a minimum lead of 2 points to secure victory. The scoring rule in a game is as follows. A score of 0 is referred to as Love, 1 point as 15, 2 points as 30, and 3 points as 40. If both players reach 3 points each, the score is called Deuce. At Deuce, the player who wins the next point gains an Advantage (either Ad-In or Ad-Out, depending on who is serving), and winning another point subsequently secures the game. If the score returns to Deuce after the Advantage, the game continues until one player wins two consecutive points.

The empirical dataset captures the performance of 32 athletes across 31 matches, comprising 37 variables and a total of 7,284 observations . In this analysis, our primary focus is on the point-level performance of Player 1. For example, in the 2023 Wimbledon final, rising young talent Carlos Alcaraz was designated as Player 1, while seasoned veteran Novak Djokovic served as Player 2. The predicted variable in our study denotes the outcome of each point for Player 1, where a value of 1 signifies a win and 0 indicates a loss.
\subsection{Verification of momentum existence}
Based on empirical data analysis, our findings indicate that extended winning and losing streaks of seven or more consecutive points occur relatively infrequently. We define $W_{7+}$ as the event of a winning streak comprising seven or more consecutive points, and $L_{7+}$ as the corresponding event for losing streaks. Analysis of all points played across 31 matches involving 32 athletes yields the contingency table presented in Table  \ref{Tab3}. 
\begin{table}[H]
    \centering
    \caption{\label{Tab3}Contingency table for the points from 31 matches}
    \begin{tabular}{|c|c|c|c|}
    \hline
    \diagbox{\textbf{Streak}}{\textbf{Subsequent point}} & \textbf{Extension} & \textbf{Termination} & $n_{i.}$ \\
    \hline
    $W_1$ & 936 & 765 & 1701 \\
    $W_2$ & 485 & 351 & 836 \\
    $W_3$ & 298 & 221 & 519 \\
    $W_4$ & 137 & 161 & 298 \\
    $W_5$ & 55 & 85 & 140 \\
    $W_6$ & 22 & 33 & 55 \\
    $W_{7+}$ & 25 & 21 & 46 \\
    \hline
    $n_{.j}$ & 1958 & 1641 & 3595 \\
    \hline
    \end{tabular}
\end{table}
Consider the null hypothesis
\[
H_0: P(\text{Extension}|W_1) = P(\text{Extension}|W_2) =  \cdots = P(\text{Extension}|W_{7+}).
\]
The statistical analysis meets all necessary assumptions for Pearson's Chi-squared test, with each expected frequency $\hat{n}_{ij} > 5$ and a sufficiently large sample size ($n = 3595 \gg 50$). The calculated test statistic yields $\chi^2 = 111.497$ with an extremely significant $p$-value of $9.51 \times 10^{-18}$. This overwhelming statistical evidence leads to the decisive rejection of the null hypothesis, strongly supporting the existence of momentum effects in men's singles tennis competitions.

\begin{figure}[h]
\centering
\includegraphics[width=0.7\linewidth]{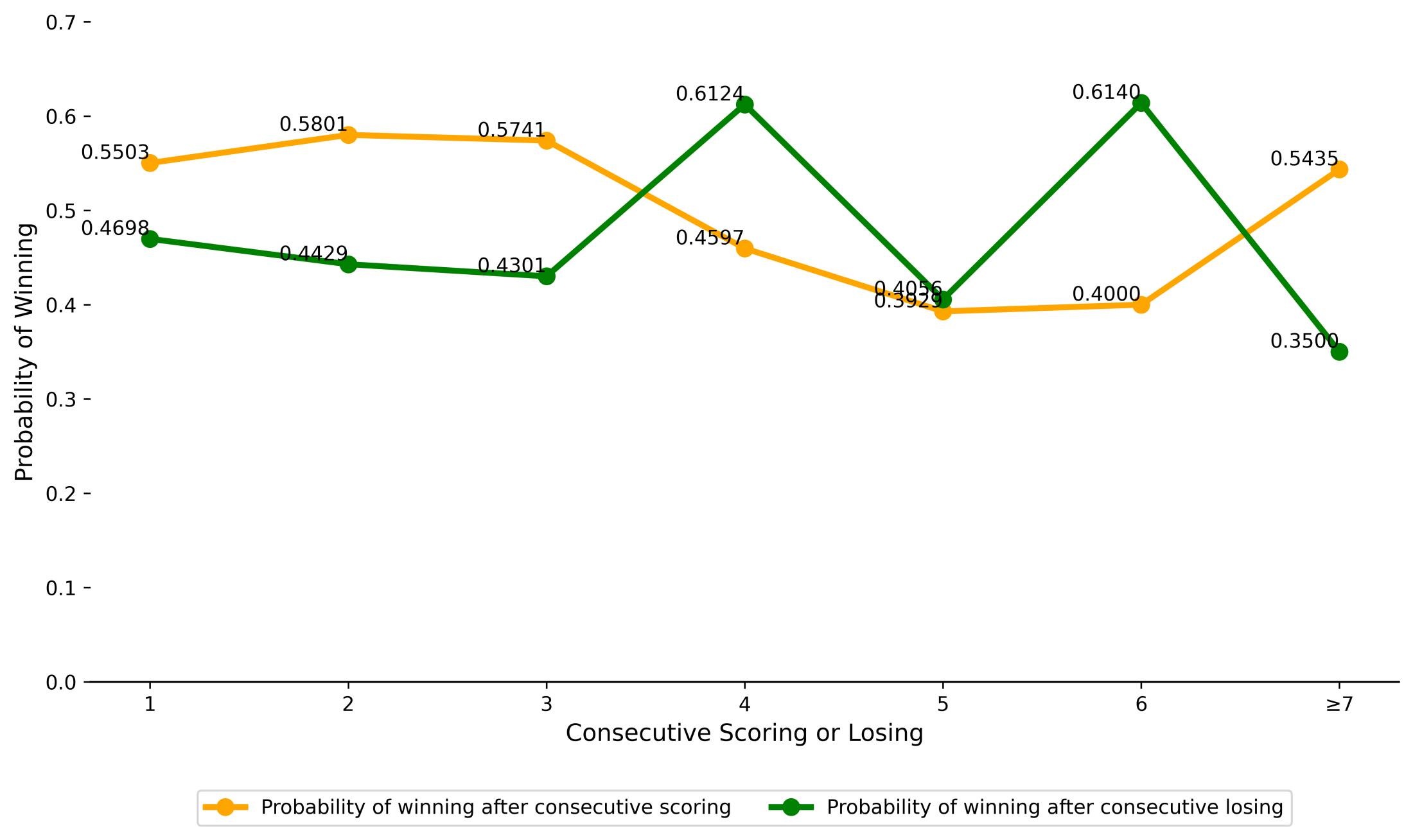}
\caption{\label{Fig2} Probability of winning after consecutive scoring or losing}
\end{figure}

In addition, Using the full sequence of point outcomes across 31 matches, we estimate the conditional probabilities of winning the subsequent point following winning or losing streaks of length $k$. These probabilities are denoted as $P(W_{\text{next}}|W_k)$ for points following a winning streak and $P(W_{\text{next}}|L_k)$ for points following a losing streak. The complete results of this analysis are systematically presented in Table \ref{Tab4} and visually summarized in Figure \ref{Fig2}.
\begin{table}[H]
    \centering
    \caption{\label{Tab4} The estimations of conditional probabilities }
    \begin{tabular}{cccc}
    \hline
         \textbf{Winning Streak} & $P(W_{\text{next}}|W_k)$ & \textbf{Losing Streak} & $P(W_{\text{next}}|L_k)$ \\
         \hline
         $W_1$ & 0.5503 & $L_1$ & 0.4698 \\
         $W_2$ & 0.5801 & $L_2$ & 0.4429 \\
         $W_3$ & 0.5741 & $L_3$ & 0.4301 \\
         $W_4$ & 0.4597 & $L_4$ & 0.6124 \\
         $W_5$ & 0.3929 & $L_5$ & 0.4056 \\
         $W_6$ & 0.4000 & $L_6$ & 0.6140 \\
         $W_{7+}$ & 0.5435 & $L_{7+}$ & 0.3500 \\
         \hline
    \end{tabular}
\end{table}

Our analysis of conditional probabilities reveals distinct momentum patterns in tennis point sequences (Figure \ref{Fig2}). For winning streaks $\{P(W_{\text{next}}|W_k)\}$, three phases emerge: initial stability ($k=1-3$) with consistent win probabilities (0.5503-0.5801), a critical transition ($k=4-6$) featuring sharp declines ($0.4597\rightarrow 0.4000$) likely due to Player 1's complacency and Player 2's adaptations, followed by equilibrium ($k \ge 7$) at 0.5435 as Player 1 establishes new performance balance. Conversely, losing streaks $\{P(W_{\text{next}}|L_k)\}$ show progressive pressure effects ($k=1-3: 0.4698\rightarrow 0.4301$), a dramatic reversal at $k=4$ (0.6124) from Player 2's relaxation or adjustments, and subsequent high volatility ($k\ge 5: 0.4056 \rightarrow 0.6140\rightarrow 0.3500$) indicating psychological fragility during prolonged adversity.

These nonlinear dynamics demonstrate that tennis momentum operates through threshold-dependent psychological mechanisms and bidirectional player adaptations. The observed phase transitions, particularly the $k=4$ inflection points in both winning and losing streaks, suggest that momentum follows complex systems principles rather than simple linear relationships. The findings reveal how players' psychological states and tactical responses interact to create emergent patterns of performance stability and volatility at different streak lengths, highlighting the nuanced nature of competitive dynamics in elite tennis.

\subsection{Measurement of the momentum }
\subsubsection{Definition of new variables}
 The empirical data includes 37 variables and the meaning of each variable is detailed in Table {\ref{tab3}}. To develop a comprehensive framework encompassing Player 1's technical prowess, physical capabilities, consistency, and strategic decision-making during competitions, we construct 16 new variables across four distinct dimensions, as elaborated below.
 \begin{table}
\centering
\caption{\label{tab3} The meaning of the original features.}
\begin{tabular}{c|c|c|c}
\hline
No. & Feature & Explanation & Example \\
\hline
1& $p1_{-}$games & games won by player 1 in current set & 0, 1, ..., 6 \\
2& $p2_{-}$games & games won by player 2 in current set & 0, 1, ..., 6 \\
3& $p1_{-}$score & player 1's score within current game & 0 , 15, 30, 40, AD  \\
4& $p2_{-}$score & player 2's score within current game & 0 , 15, 30, 40, AD  \\
5& server & server of the point & 1: player 1, 2: player 2 \\
6& serve no & first or second serve & 1: first serve, 2: second serve \\
7& point victor & winner of the point & 1, 2 \\
8& $p1_{-}$points$_{-}$won & number of points won by player 1 in match & 0, 1, 2... etc. \\
9& $p2_{-}$points$_{-}$won & number of points won by player 2 in match & 0, 1, 2... etc. \\
10& game victor & a player won a game this point & 0: no one, 1: player 1, 2: player 2 \\
11& set victor & a player won a set this point & 0: no one, 1: player 1, 2: player 2 \\
12&  $p1_{-}$ace & player 1 hit an untouchable winning serve & 0 or 1 \\
13& $p2_{-}$ace & player 2 hit an untouchable winning serve & 0 or 1 \\
14& $p1_{-}$winner & player 1 hit an untouchable winning shot & 0 or 1 \\
15& $p2_{-}$winner & player 2 hit an untouchable winning shot & 0 or 1 \\
16& winner shot type & category of untouchable shot & F: Forehand, B: Backhand \\
17& $p1_{-}$double$_{-}$fault & player 1 missed both serves and lost the point & 0 or 1 \\
18& $p2_{-}$double$_{-}$fault & player 2 missed both serves and lost the point & 0 or 1 \\
19& $p1_{-}$unf$_{-}$err & player 1 made an unforced error & 0 or 1 \\
20& $p2_{-}$unf$_{-}$err & player 2 made an unforced error & 0 or 1 \\
21& $p1_{-}$net$_{-}$pt & player 1 made it to the net & 0 or 1 \\
22 &$p2_{-}$net$_{-}$pt & player 2 made it to the net & 0 or 1 \\
23&$p1_{-}$net$_{-}$pt$_{-}$won & player 1 won the point while at the net & 0 or 1 \\
24&$p2_{-}$net$_{-}$pt$_{-}$won & player 2 won the point while at the net & 0 or 1 \\
25&$p1_{-}$break$_{-}$pt & Break point for Player 1 & 0 or 1 \\
26&$p2_{-}$break$_{-}$pt & Break point for Player 2 & 0 or 1 \\
27&$p1_{-}$break$_{-}$pt$_{-}$won & player 1 won a game by player 2's serve & 0 or 1 \\
28&$p2_{-}$break$_{-}$pt$_{-}$won & player 2 won a game by player 1's serve & 0 or 1 \\
29&$p1_{-}$force$_{-}$err & player 1 made a forced error & 0 or 1 \\
30&$p2_{-}$force$_{-}$err & player 2 made a forced error & 0 or 1 \\
31&ball$_{-}$speed & speed of the ball in km/h & 5.376, 21.384, etc. \\
32&ball$_{-}$spin & spin rate of the ball in rpm & 6.485, 12.473, etc. \\
33&rally$_{-}$length & number of shots in the point & 1, 2, 4, etc. (includes serve) \\
34&game$_{-}$time & time taken to complete the game in seconds & 81, 124, etc. \\
35&serve$_{-}$direction & direction of the serve & B, BC, BW, , etc. \\
36&serve$_{-}$depth & how close the serve lands to the service line & CTL, NCTL \\
37&return$_{-}$depth & how deep the return lands in the court & D: Deep, ND: Not Deep \\
\hline
\end{tabular}
\end{table}

 \begin{itemize}
    \item \textbf{Game Outcome and Score Dynamics}  
    This dimension captures the overall results and score fluctuations throughout the match, providing insight into the competitive position at any given time. The following variables are included.
    \begin{itemize}
        \item $x_1$: Number of games won by Player 1
        \item $x_2$: Score difference between Player 1 and Player 2
        \item $x_3$: Whether Player 1 is serving
        \item $x_4$: Player 1’s lead status
    \end{itemize}

    \item \textbf{Technical Skills and Performance}  
    This dimension focuses on the technical aspects of the match, including serving and hitting key shots such as winners and aces. The following variables are included.
    \begin{itemize}
        \item $x_5$: First serve attempt
        \item $x_6$: Number of aces by Player 1
        \item $x_7$: Number of winning shots by Player 1
        \item $x_8$: Net points won ratio by Player 1
        \item $x_9$: Break points won ratio by Player 1
    \end{itemize}

    \item \textbf{Stability and Error Control}  
    This dimension reflects Player 1's consistency and mental stability throughout the match by focusing on errors and faults. The following variables are included.
    \begin{itemize}
        \item $x_{10}$: Double faults committed by Player 1
        \item $x_{11}$: Unforced errors committed by Player 1
    \end{itemize}

    \item \textbf{Physical Effort and Shot Intensity}  
    This dimension assesses the physical demands on Player 1 during the match, as well as the intensity of the shots they hit, which can be crucial in determining the outcome of the match. The following variables are included.
    \begin{itemize}
        \item $x_{12}$: Total distance run by Player 1
        \item $x_{13}$: Distance run by Player 1 in the last three points
        \item $x_{14}$: Distance run by Player 1 during the current point
        \item $x_{15}$: Ball speed during the current point
        \item $x_{16}$: Ball speed multiplied by the number of serves
    \end{itemize}
\end{itemize}

\begin{table}[H]
\centering
\caption{\label{tab4} Definitions of new features}
\begin{tabular}{c|c|c}
\hline
Variable & Definition & Formula/Source \\
\hline
$x_1$ & Matches won by Player 1 & Games won by Player 1 in current set \\
$x_2$ & Score difference between players & Player 1’s score - Player 2’s score \\
$x_3$ & Whether it’s Player 1’s first serve & 1 if first serve, else 0 \\
$x_4$ & Player 1’s lead status & 1 if Player 1’s score $\geq$ Player 2’s, else 0 \\
$x_5$ & Set difference between players & Player 1’s sets won - Player 2’s sets won \\
$x_6$ & Player 1 scored an ace & 1 if Player 1 hits an ace, else 0 \\
$x_7$ & Player 1 scored a winning shot & 1 if Player 1 hits a winner, else 0 \\
$x_8$ & Double faults by Player 1 & 1 if Player 1 commits a double fault, else 0 \\
$x_9$ & Unforced errors by Player 1 & 1 if Player 1 makes an unforced error, else 0 \\
$x_{10}$ & Net points won ratio by Player 1 & Points won at net / Total net points \\
$x_{11}$ & Break points won ratio by Player 1 & Break points won / Total break points \\
$x_{12}$ & Total distance run by Player 1 & Cumulative distance run by Player 1 \\
$x_{13}$ & Distance run in last three points & Distance run in last three points \\
$x_{14}$ & Distance run during current point & Distance run during the current point \\
$x_{15}$ & Ball speed during current point & Speed of the ball during current point \\
$x_{16}$ & Ball speed $\times$ number of serves & Ball speed $\times$ number of serves \\
\hline
\end{tabular}
\end{table}

\subsubsection{Feature selection based on stepwise logistic regression with AUC criterion}
\begin{figure}[H]
\centering
\includegraphics[width=0.8\linewidth]{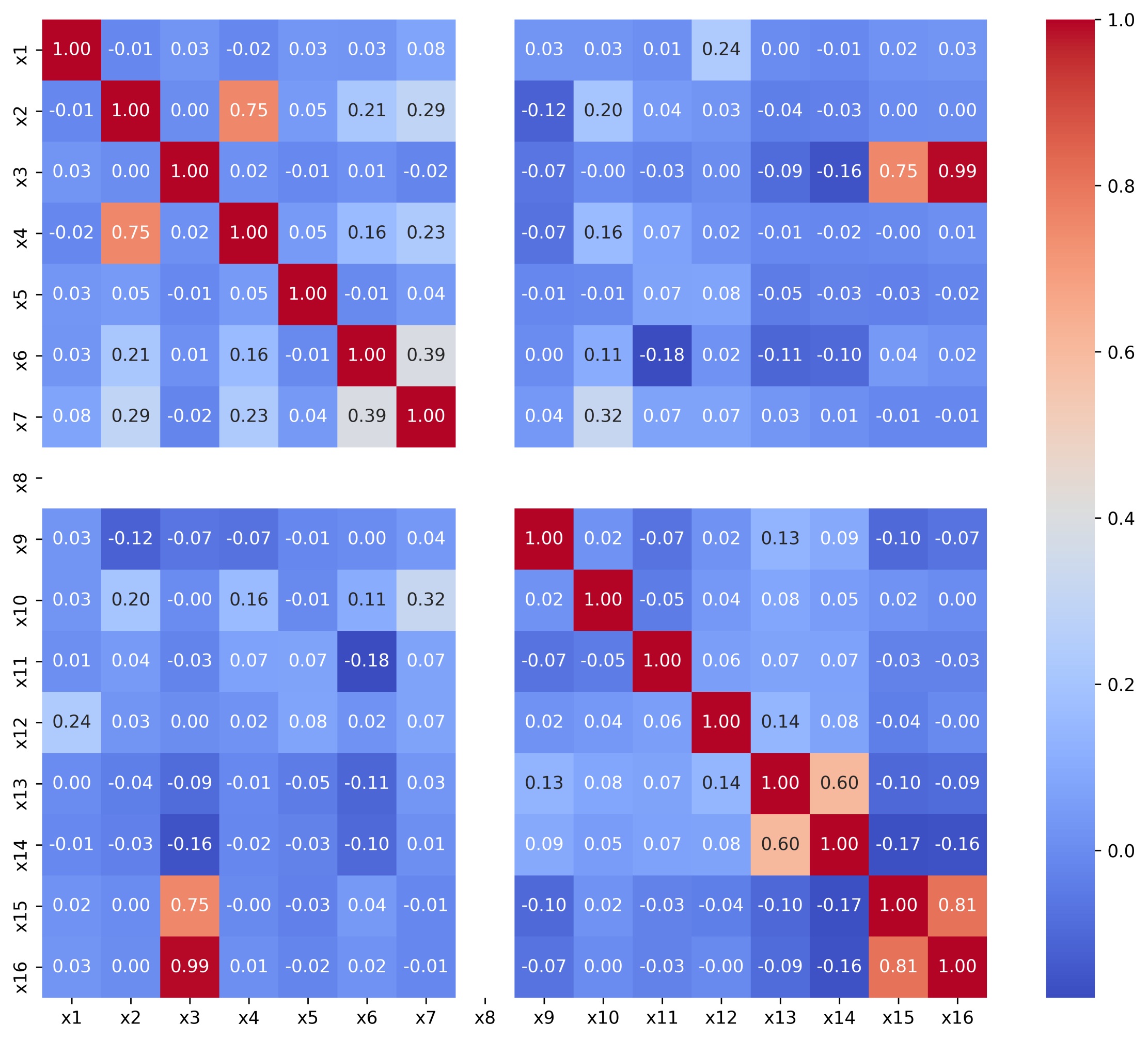}
\caption{Feature correlation heatmap}
\label{fig:correlation_heatmap}
\end{figure}

To ensure the interpretability of the momentum measurement, it is necessary to select the important features from the 16 new variables in Table \ref{tab4}. Additionally, feature selection can also overcome the issue of multicollinearity that undermines the robustness and generalization ability of the predictive model.
Before making feature selection, we conduct a correlation analysis of the 16 features.
Figure \ref{fig:correlation_heatmap} is the Pearson correlation heatmap, where the colour intensity represents the degree of correlation between two features. 

As can be seen from Figure \ref{fig:correlation_heatmap}, there is a significant linear correlation between some of the variables. For example, the variable $x_3$ (Whether it’s Player 1’s first serve ) is highly correlated with $x_{16}$ (ball speed multiplied by the number of serves), with a correlation coefficient of 0.99. It means that $x_{16}$ can be defined by $x_3$, either of them could be removed during feature selection. Additionally, a moderately strong positive correlation ($\rho = 0.75$) exists between $x_2$ (score difference between Player 1 and Player 2) and $x_{15}$ (ball speed during the current point), suggesting that a faster serve correlates with a higher score difference in favor of Player 1. This relationship indicates some redundancy, though both features provide valuable insights and should be carefully considered. Similarly, $x_{13}$ (distance run by Player 1 in the last three points) and $x_{14}$ (distance run by Player 1 during the current point) show moderate correlation ($\rho = 0.60$), which may reflect the increased physical effort during critical break points, potentially influencing match outcomes.

In contrast, the correlation coefficients of $x_8$ and other features with themselves are not shown, because the majority of the observations for variable $x_8$ are zero, resulting in a blank space in the heatmap. Due to its sparsity, it is a potential candidate for elimination. The low correlation between most of the other feature pairs indicates that they provide independent information, reducing the risk of multicollinearity, and ensuring that each feature contributes unique insights into the prediction of match results. This independence benefits the performance of the model and ensures that redundant features do not distort the predictive power of the model.

We take the outcome (winning or losing) of Player 1 in each point  as the target variable, while the 16 features listed in Table \ref{tab4} are treated as original predictors. The logistic regression model, utilizing the area under the curve (AUC) of receiver operating characteristic (ROC) as its criterion, is employed for feature selection. In other words, we aim to select an optimal subset of features that maximizes the AUC of the model. Resorting to the idea of stepwise selection , the process of feature selection can be carried as follows.
\begin{itemize}
    \item  Begin by considering the model with no predictors.
    \item  Sequentially add the most statistically significant feature to the model, one at a time, based on AUC criterion.
    \item  After adding each predictor, re-evaluate the significance of all predictors currently in the model.If the removal of a certain variable results in an increase in the AUC value, then that variable should be deleted from the existing feature subset.
    \item  Continue this process until neither new significant features can be added nor insignificant variables can be removed, that is , the value of AUC does not increase or decreases.
\end{itemize}
 
Initially, logistic models are built for each of the 16 features in sequence, and it shows that the AUC corresponding to the feature $x_7$ reached a maximum of 0.6415.
Subsequently, we apply the above stepwise algorithm to complete feature selection.
Then the features $x_6, x_{10}, x_4, x_9$ and $x_3$ are sequentially selected into the models with a maximum value of AUC 0.7275. The process is outlined in Table \ref{tab5}.
\begin{table}[H]
\centering
\caption{\label{tab5} The feature selection process}
\begin{tabular}{c|c|c}
\hline
Model & Features & AUC \\
\hline
Model 1 & $x_7$ & 0.6415\\
Model 2 & $x_7,x_6$ & 0.6786\\
Model 3 & $x_7,x_6,x_{10}$ & 0.7014\\
Model 4 & $x_7,x_6,x_{10},x_4$ & 0.7079\\
Model 5 & $x_7,x_6,x_{10},x_4,x_9$ & 0.7215\\
Model 6 & $x_7,x_6,x_{10},x_4,x_9,x_3$ & 0.7275\\
\hline
\end{tabular}
\end{table}

\subsubsection{Measurement of the momentum }
Taking the data from the final match as an example, and applying the entropy weight method introduced in Section 2.2, we can derive the weight of each selected feature, as shown in Table  \ref{tab7}.

\begin{table}[H]
\centering
\caption{\label{tab7}The weights of selected features}
\begin{tabular}{c|ccccccc}
\hline
\textbf{Feature} & $z_3$ & $z_4$ & $z_6$ & $z_7$ & $z_9$ & $z_{10}$ \\
\hline
\textbf{Weight} & 0.3959 & 0.1122 & 0.0987 & 0.1162 &  0.1136 & 0.1633 \\
\hline
\end{tabular}
\end{table}

Then we get the measurement of momentum  as follows.
\begin{equation}\label{e1}
 M_t = 0.3959z_{3t} + 0.1122z_{4t} + 0.0987z_{6t} + 0.1162z_{7t} + 0.1136z_{9t} + 0.1633z_{10t},
\end{equation}
 where $z_{it}$ is the standardization of the feature $x_{it}$.

\begin{figure}[H]
\centering
\includegraphics[width=0.8\textwidth]{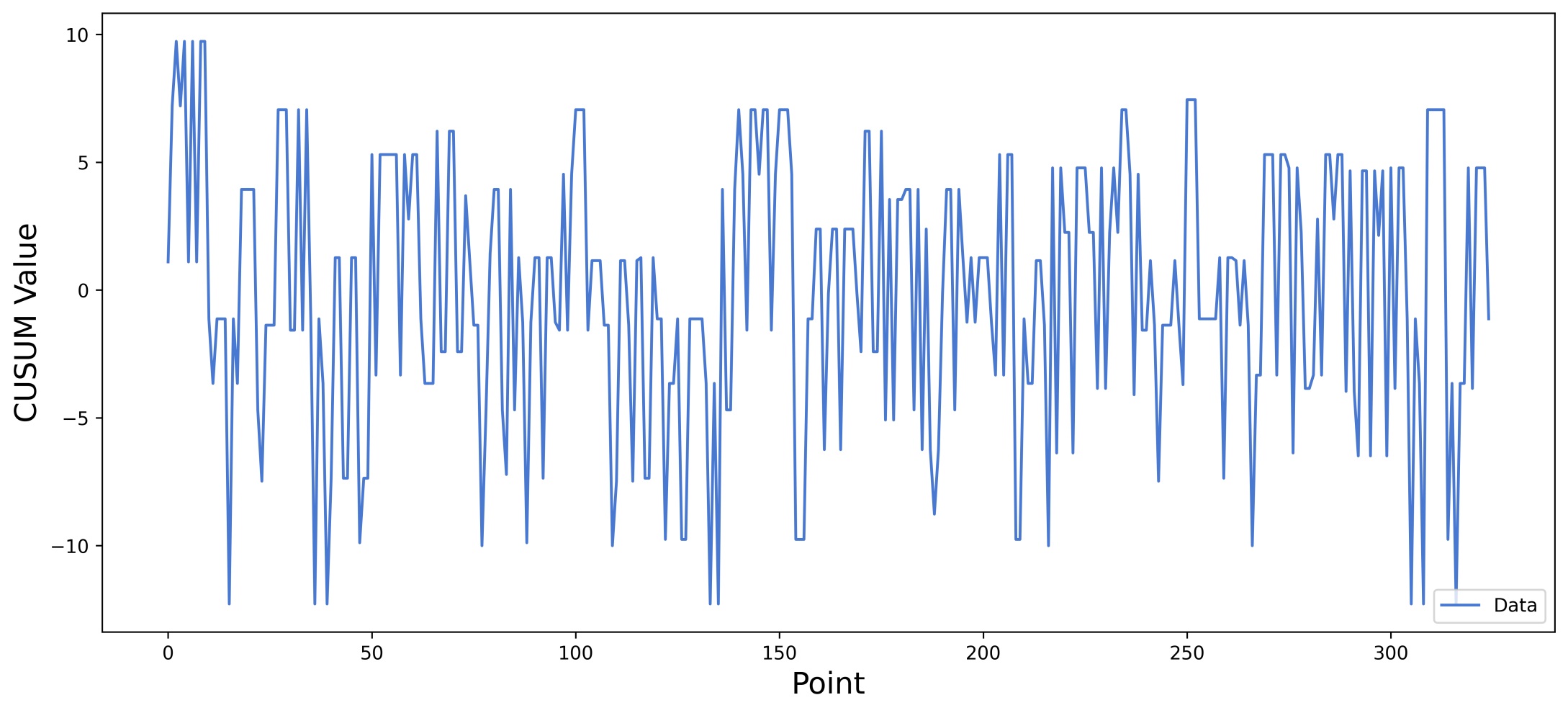}
\caption{Time series chart of $M_t$ for Carlos Alcaraz in the final match.}
\label{fig:TimeSeries_Mt}
\end{figure}

 By equation (\ref{e1}), we calculate the momentum $M_t$ for Carlos Alcaraz at each point in the final match.
 Figure \ref{fig:TimeSeries_Mt} presents the time series chart of $M_t$ for 325 points.

As shown in Figure \ref{fig:TimeSeries_Mt}, there exist two obvious characteristics as follows.

\begin{itemize}
    \item \textbf{Frequent Fluctuations.} The momentum curve exhibits frequent fluctuations, characterized by regular upward and downward movements. These variations signify shifts in the psychological and physical states of the players, often aligning with pivotal events such as streaks of consecutive points won or lost.
    
    \item \textbf{Potential Change Points.} Within the momentum curve, there are visible sections where rapid and substantial shifts occur. These sudden alterations hint at crucial turning points in the match, marking moments where momentum could swing dramatically in favor of one player over the other.
        \end{itemize}

Although visual inspection provides an intuitive sense of momentum changes, we need to provide the interpretation through statistical validation.
Next, we will utilize cumulative sum (CUSUM) control chart to identify change points in $M_t$.

\subsection{Change point detection by CUSUM method}
To ensure a consistent and interpretable number of change points, we establish a target number of change points and employ an algorithm that dynamically adjusts the threshold h to achieve this target. If the number of detected change points exceeds the target, the algorithm increases \( h \) by 10\% to decrease sensitivity. Conversely, if the number of detected change points is less than the target, the algorithm decreases \( h \) by 10\% to increase sensitivity. The iterative adjustment process continues until the total number of detected change points converges to the target value  within a tolerance of 1\%. This dynamic adjustment mechanism ensures that the CUSUM algorithm identifies a stable number of change points, balancing between sensitivity and reliability.

\begin{figure}[H]
\centering
\includegraphics[width=0.8\textwidth]{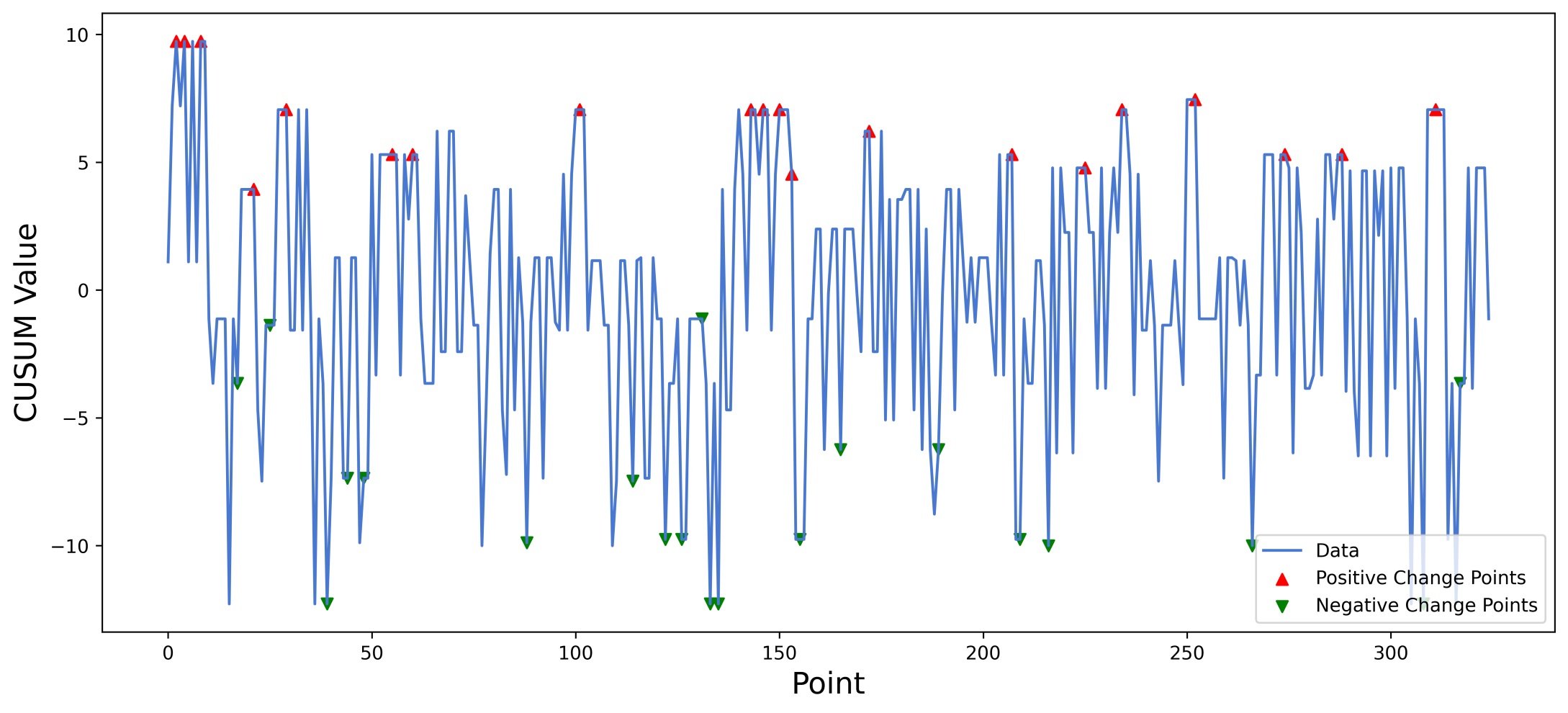}
\caption{Detection results of change points of $M_t$ in the final match}
\label{fig:cusum_detection}
\end{figure}

For the momentum time series $M_t$ in the final match, 40 change points are detected, including 20 positive change points (indicating improved performance) and 20 negative change points (indicating decreased performance). Figure \ref{fig:cusum_detection} shows the distribution of change points in the time series $M_t$, where red triangles correspond to positive change points and green triangles correspond to negative change points. 

\begin{figure}[H]
\centering
\includegraphics[width=0.8\textwidth]{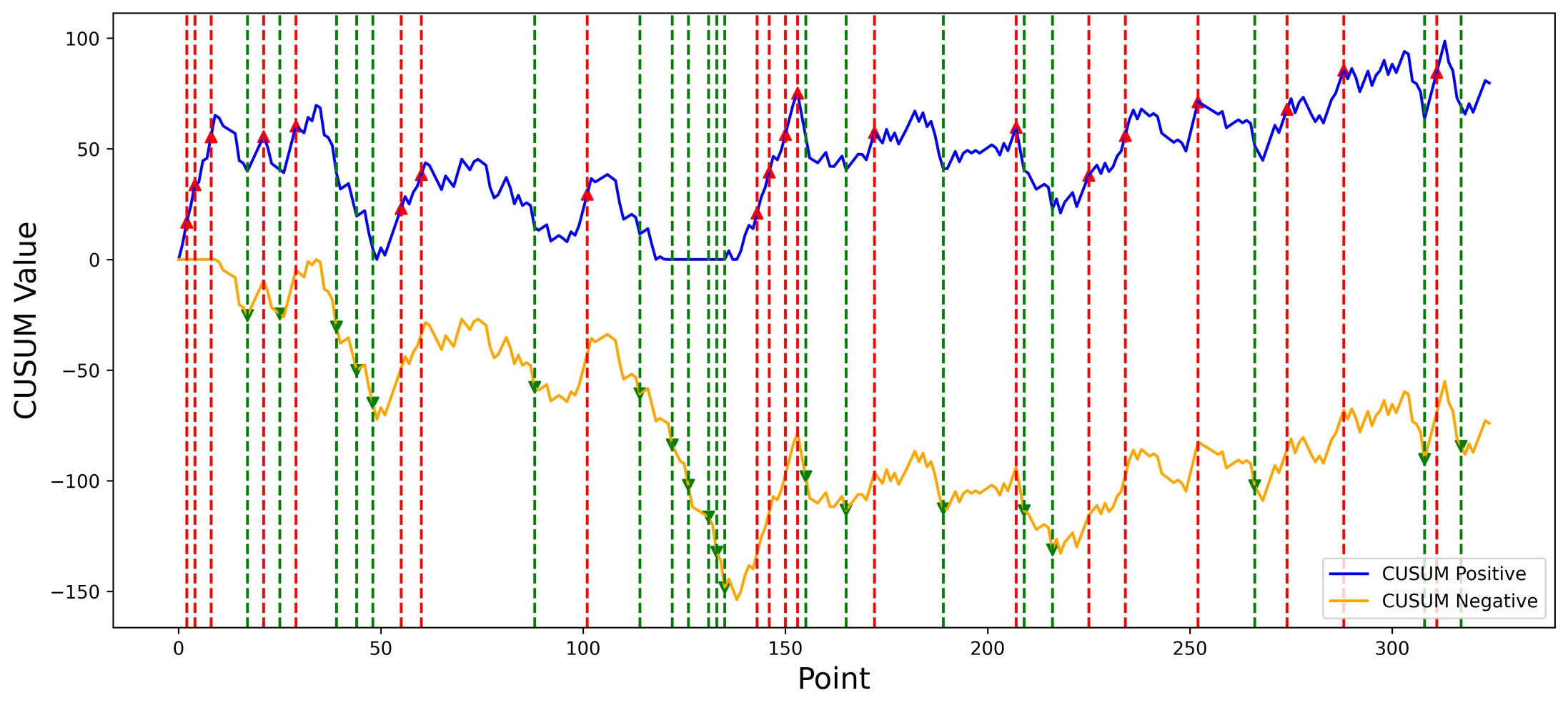}
\caption{CUSUM curves for change points of $M_t$ in the final match}
\label{fig:cusum_analysis}
\end{figure}

The alternating pattern of positive and negative change points reveals frequent shifts in momentum advantage between the two players, indicating a dynamic and varied competitive landscape with changing control over the match. To further analyze the distribution and impact of change points, the identified positive and negative change points are superimposed on the CUSUM curves.  See Figure \ref{fig:cusum_analysis}. It is evident that positive change points generally coincide with sharp increases in the positive CUSUM values, signaling a robust upward shift in the player's  momentum. Conversely, negative change points align with steep declines in the negative CUSUM values, indicating a substantial loss of momentum. The results demonstrate that the CUSUM method can efficiently captures dynamic shifts in the match's momentum and offers insightful understanding into the variations in performance over time.

\begin{figure}[H]
\centering
\includegraphics[width=0.8\textwidth]{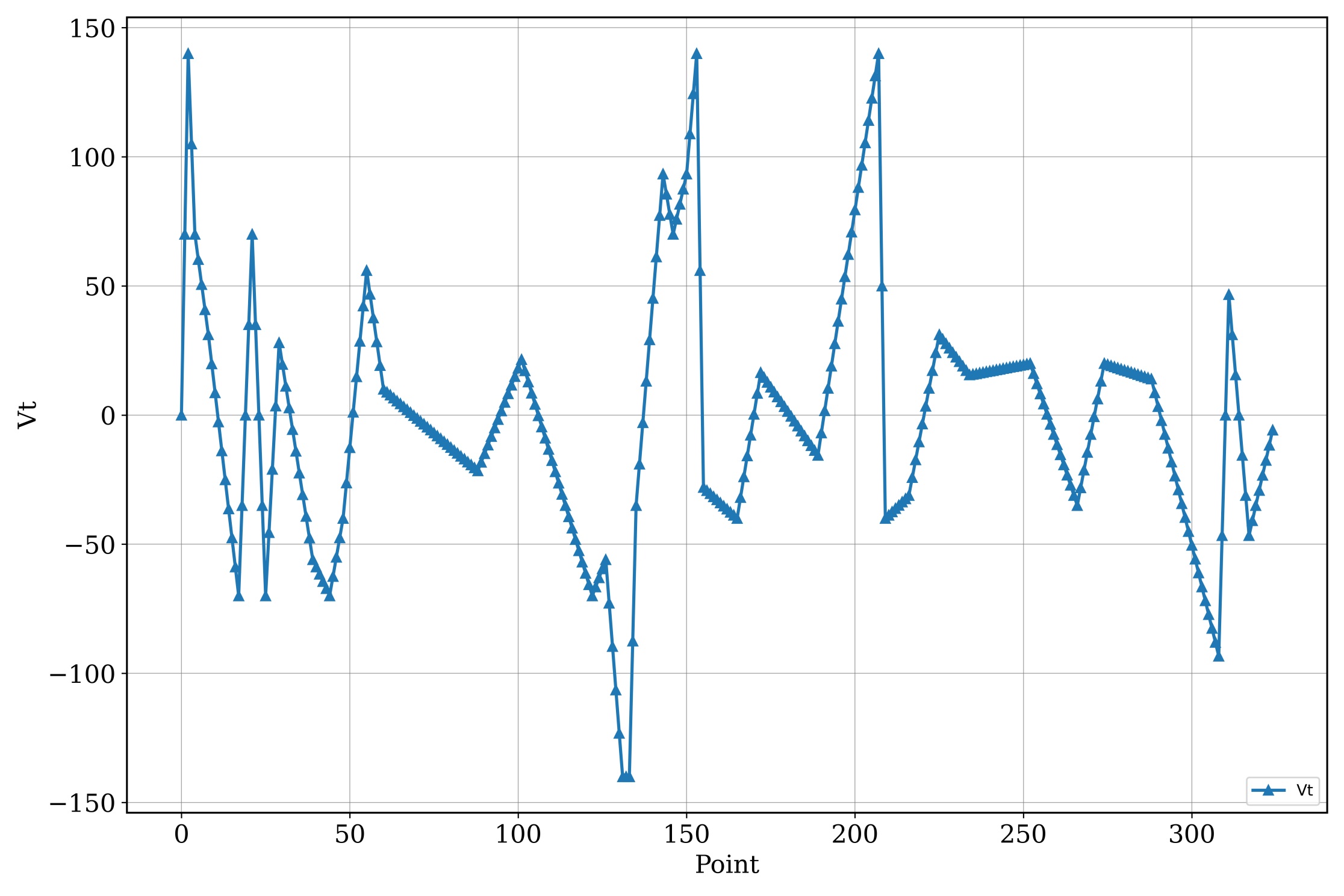}
\caption{Intensity of momentum shifts in the final match}
\label{fig:Vt_analysis}
\end{figure}

To analyze the intensity of momentum shifts, we calculate the relative distance \(V_t\) according to the definition in Section 2.4. See Figure \ref{fig:Vt_analysis}.

The series $V_t$ exhibits significant fluctuations throughout the course of the final match. The sharp spikes in the graph represent the intensity of momentum shifts at change points. Positive peaks ($V_t>0$) indicate strong momentum shifts in favor of Carlos Alcaraz at these change points, with larger peaks corresponding to higher shift intensities. Conversely, negative peaks ($V_t<0$) denote strong momentum shifts against Carlos Alcaraz (i.e., momentum loss or momentum gained by Novak Djokovic) at these change points, with greater absolute values indicating higher intensities. Between two adjacent change points, the value of $V_t$ is obtained through linear interpolation, reflecting a smooth transition of momentum intensity within these intervals. Before the first change point, $V_t$ increases linearly from 0 to the intensity value at the first change point. After the last change point, $V_t$ decreases linearly, reflecting the gradual weakening of momentum effects as the match approaches its conclusion, ultimately tending towards 0.

This sequence graph of $V_t$ visually reveals the dynamic characteristics of momentum shift intensities during the match, identifying not only the moments when significant momentum changes occur but also quantifying the severity of these changes. It provides a crucial quantitative indicator for incorporating momentum effects into subsequent predictive models.

 \subsection{Prediction based on BP+PSO model }
\begin{figure}[H]
\centering
\includegraphics[width=0.8\textwidth]{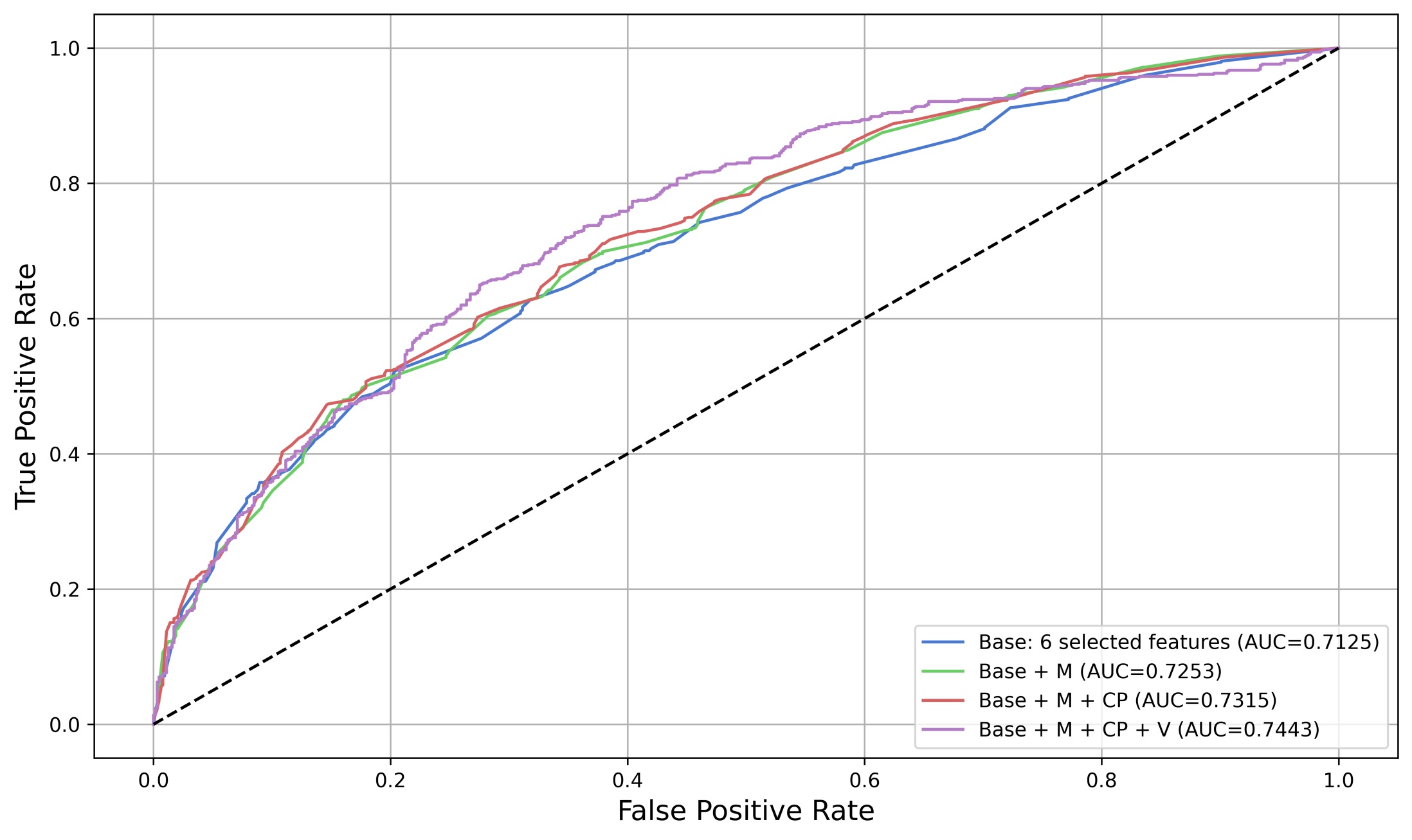}
\caption{ROC curve comparison under four input layers.}
\label{fig:ROC1}
\end{figure}

 The BP+PSO model is initially trained using the data of final match to determine the optimal configuration of network weights and nodes. This refined configuration is then validated on the data of all 31 matches to assess the model's performance on a larger scale and evaluate its ability to generalize across diverse match scenarios.
Denote the six selected features as a vector
\(Base=(X_{3}, X_{4}, X_{6}, X_{7}, X_{9}, X_{10})\).\\
 The data is divided into training set and test set in an 8:2 ratio. Table \ref{RFIL} and Figure \ref{fig:ROC1} present the prediction evaluation metrics and ROC curve on the test set for four input layers, respectively.

\begin{table}[H]
\centering
\caption{ \label{RFIL} Prediction performance of BP+PSO model under four input layers}
\begin{tabular}{|c|c|c|c|c|}
\hline
Input layer & Precision & Recall & F1-score & AUC \\ \hline
 \(Base\) & 0.6479 & 0.6855 & 0.6661 & 0.7125 \\ \hline
 \(Base+ M\) & 0.6610 & 0.6945 & 0.6773 & 0.7253 \\ \hline
 \(Base+ M + CP\) & 0.6643 & 0.6990 & 0.6812 & 0.7315 \\ \hline
 \(Base+ M + CP+ V\) & 0.6963 & 0.6766 & 0.6863 & 0.7443 \\ \hline
  \end{tabular}
\end{table}

As shown in Table~\ref{RFIL}, the inclusion of momentum-related features significantly improves the prediction performance of the BP+PSO model. When using only the selected base features $Base$, the model achieves an AUC of 0.125. By adding the momentum measure $M$, the AUC increases to 0.7253, and the F1 score improves from 0.6661 to 0.6773, demonstrating the effectiveness of $M$ in enhancing predictive accuracy. When the change point label $CP$ is added, the AUC further improves to 0.7315, though recall drops slightly. Finally, the inclusion of the momentum intensity $V$ leads to the best overall performance, achieving the highest AUC of 0.7443, along with a precision of 0.6963. It suggests that $V$ provides additional and important information in predicting the outcome of competition.

\begin{figure}[H]
\centering
\includegraphics[width=0.8\textwidth]{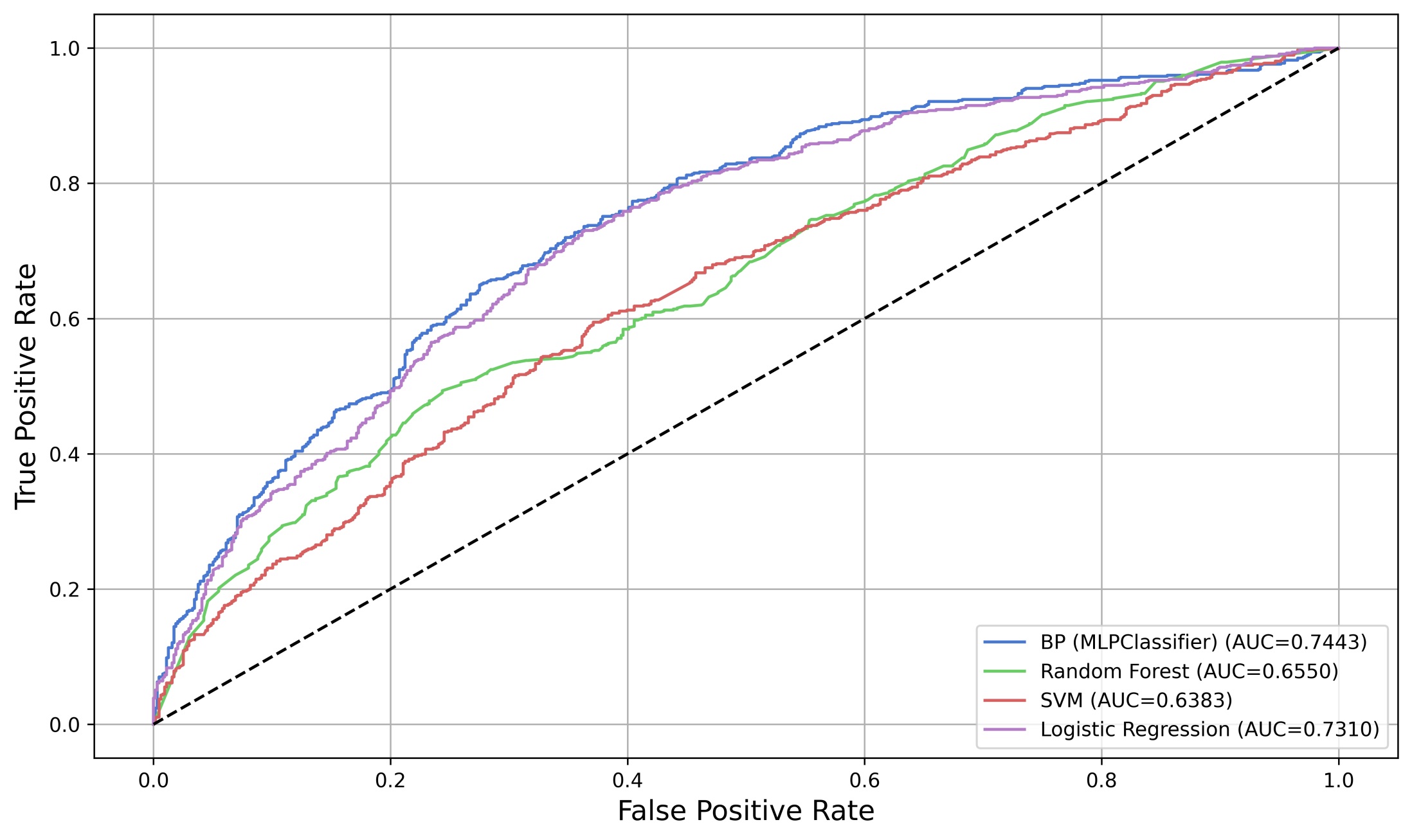}
\caption{ROC curve comparison of four models.}
\label{fig:ROC2}
\end{figure}

In addition to the BP+PSO model, we further evaluate three classical machine learning algorithms—Random Forest, Support Vector Machine (SVM), and Logistic Regression—under the same input layer (\(Base + M + CP + V\)). The prediction results of all models are summarized in Table~\ref{Other models} and Figure ~\ref{fig:ROC2}.
\begin{table}[H]
\centering
\caption{\label{Other models} Prediction performance of four  machine learning algorithms}
\begin{tabular}{|c|c|c|c|c|}
\hline
Feature Combination & Precision & Recall & F1-score & AUC \\ \hline
Random Forest  & 0.6052 & 0.6213 & 0.6131 & 0.6550 \\ \hline
SVM  & 0.5881 & 0.7012 & 0.6397 & 0.6383 \\ \hline
Logistic Regression  & 0.6886 & 0.6967 & 0.6926 & 0.7310 \\ \hline
BP+PSO &0.6963 & 0.6766 &0.6863 &0.7443\\ \hline
\end{tabular}
\end{table}

As shown in Table~\ref{Other models}, the BP+PSO model achieves the highest precision (0.6963) and AUC (0.7443), confirming its superiority in both classification accuracy and overall discrimination. Logistic Regression also performs relatively well across all metrics. Random Forest ranks last in terms of Recall and F1-score, and comes in second-to-last place for precision and AUC. On the other hand, SVM performs worst in precision and AUC, and takes the second-to-last position for recall and F1-score. while SVM shows strong recall but suffers from lower precision and AUC. Random Forest performs not well in four , indicating limited effectiveness under the current feature configuration.
\subsection{Importance analysis of features based on SHAP }
\begin{figure}[H]
\centering
\includegraphics[width=0.8\textwidth]{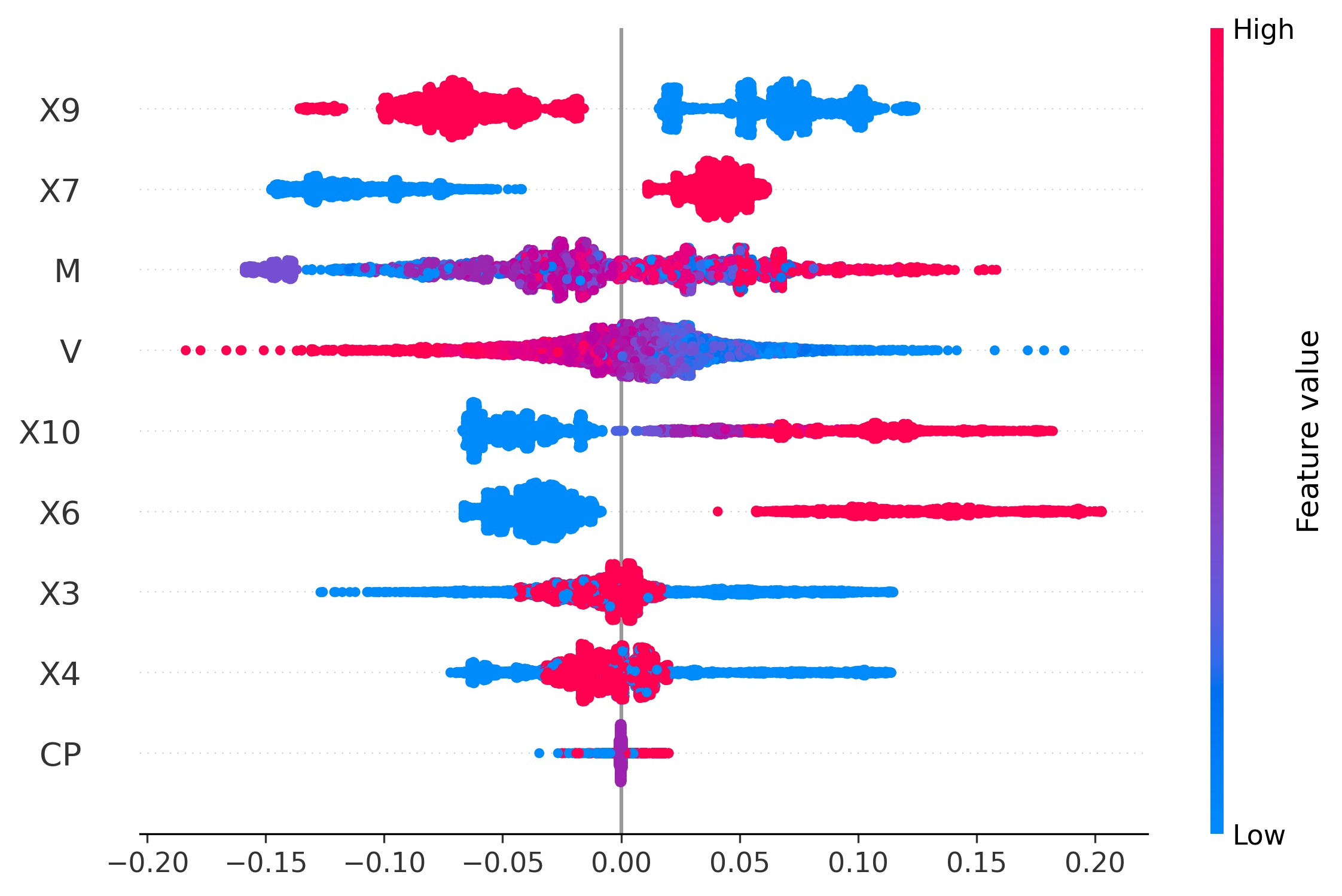}
\caption{
SHAP values of input features under BP+PSO model.}
\label{fig:SHAP}
\end{figure}

\begin{figure}[H]
\centering
\includegraphics[width=0.8\textwidth]{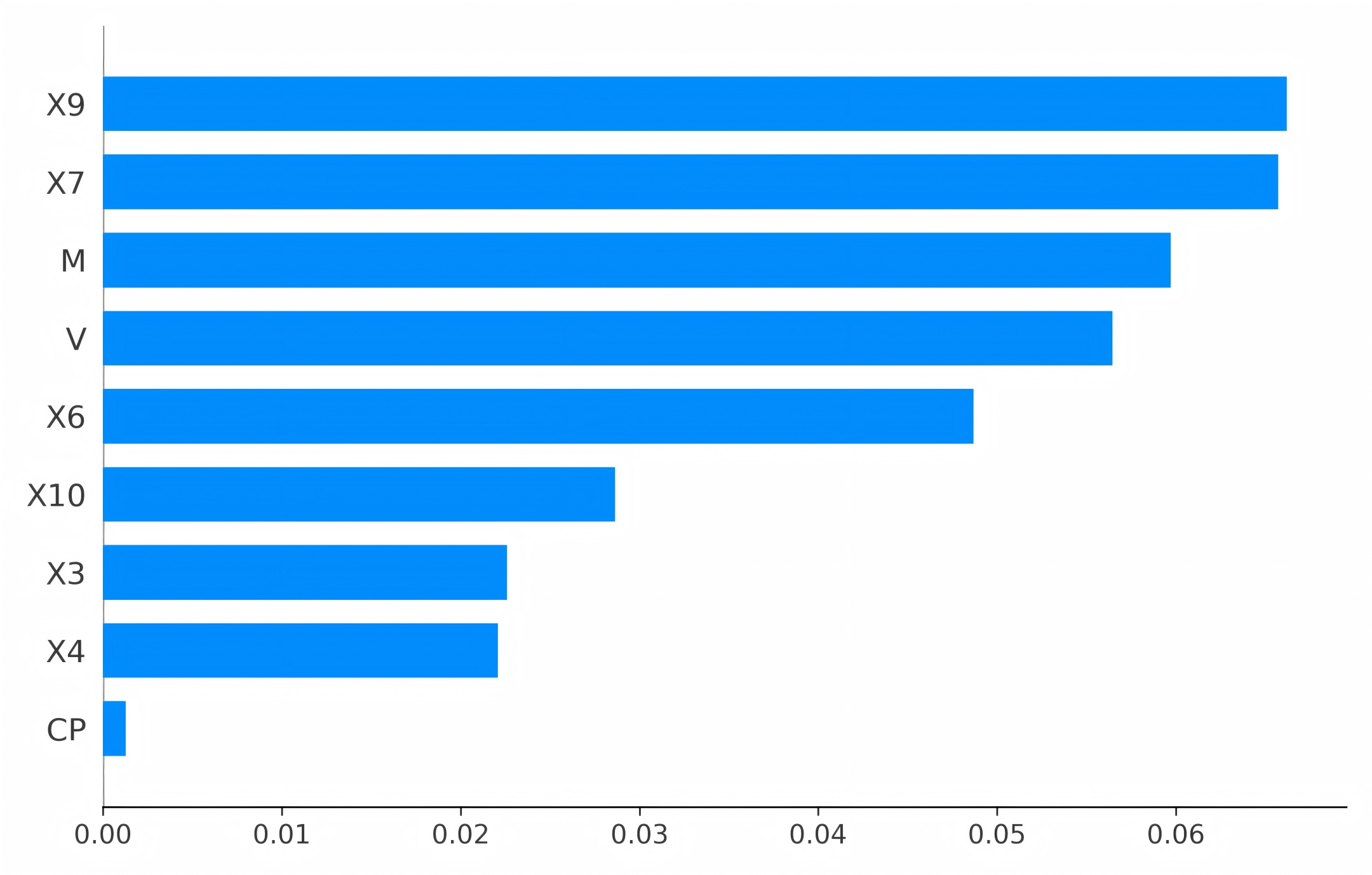}
\caption{Mean absolute of SHAP values of input features under BP+PSO model.}
\label{fig:MeanSHAP}
\end{figure}

Based on the BP+PSO model, we obtain the SHAP values and the mean absolute SHAP values for each feature in the input layer. These results are illustrated in Figure \ref{fig:SHAP} and Figure \ref{fig:MeanSHAP}. A detailed examination reveals that the feature $X_9$ (Player 1's unforced error) emerges as the most influential feature in predicting the outcome of a point, exerting a strong negative impact. Specifically, when Player 1 commits an unforced error at the current point, the SHAP value becomes significantly negative, thereby substantially reducing the probability of winning that point. This finding underscores the fundamental importance of minimizing unforced errors in a match. Even if a player cannot execute particularly spectacular winning shots, consistently returning the ball within the court boundaries and forcing the opponent to exert effort can enhance their chances of winning the point. The coach's oft-repeated mantra, "making fewer mistakes means winning," is thus empirically validated. On crucial points, the mindset and ability to "not give away points" become particularly vital, as they directly deprive the opponent of easy scoring opportunities.

The feature $X_7$ (Player 1's winning shot) ranks closely behind $X_9$ and exhibits a strong positive impact. When Player 1 hits a winning shot at the current point, the SHAP value becomes significantly positive, thereby substantially increasing the likelihood of winning the point. If $X_9$ represents the defensive baseline, then $X_7$ embodies the offensive blade. When an athlete can execute a winning shot that the opponent cannot reach, they directly secure the point. This not only highlights the athlete's shot quality and power but also demoralizes the opponent and reinforces their dominant position in the rally. During intense rallies, a high-quality winning shot can often break the deadlock instantly and provide a decisive advantage.

It is noteworthy that the two features of momentum effect, $M$ and $V$, also play a relatively significant role in predicting the outcome of a point, ranking third and fourth, respectively. The momentum $M$ has a strong positive impact on the prediction results, whereas the intensity of momentum transfer $V$ exerts a strong negative influence. When Player 1 possesses greater momentum at the current point, it indicates an increased likelihood of winning that point. Conversely, when the relative distance $V$ for Player 1 at the current point is higher, it suggests a stronger intensity of momentum transfer, which in turn reduces Player 1's probability of winning the point.

Regarding the feature $X_{10}$ (Net points won ratio by Player 1 in the current game), although it is not as crucial as $X_9$ and $X_7$
, its positive correlation remains significant. When Player 1 achieves a high net points won ratio in the current game, the SHAP value tends to be positive, indicating an increased probability of winning the current point. When a player can consistently score points at the net and maintain a high net points won ratio, it signifies successful execution of net play tactics, forcing the opponent to alter their strategy and effectively shortening the rally. This tactical success provides the player with a psychological edge and establishes a positive momentum within the game. Even if the current point is not a net play, previous successful net play experiences can boost the player's confidence and put the opponent on the defensive, thereby enhancing their overall probability of winning subsequent points.

For the remaining four features, their relative importance is comparatively lower. The change point label $CP$ ranks last, as its information is already encapsulated within the feature $V$.

\section{Conclusions and discussion}

In this paper, we propose a novel framework to study the momentum effect in singles tennis and do an empirical study based on the data from 2023 Wimbledon Men’s Singles
competition. Firstly, we verify the existence of momentum in men's singles tennis matches by applying Pearson Chi-squared test and conditional probability analysis through a frequency analysis of winning after consecutive scorings. Secondly, based on feature selection using logistic stepwise regression combined with AUC criterion, we apply the entropy weight method to construct a momentum measurement indicator \(M\). Thirdly, we discuss the identification of change points in the time series (\(M\)) using CUSUM control chart and provide a measurement of momentum transfer intensity by defining the concept of relative distance \(V\). Finally, we establish prediction models using the BP neural network combined with PSO algorithm. The results show that the  momentum measurement indicator \(M\) and momentum transfer intensity indicator \(V\) constructed in this paper can provide important additional information for predicting the outcome of the point and significantly improving the prediction performance. Furthermore, compared to other machine learning algorithms such as Random Forest, SVM, and logistic regression, the BP+PSO model demonstrates a clear competitive advantage.

In the future, we aim to refine the prediction of athletes' competition outcomes by focusing on two key areas: feature selection and the application of prediction models. We can improve the measurement of momentum by selecting important features from three dimensions: psychological, technical, and physical, combined with statistical methods. As for the application of prediction models, the empirical results indicate that different machine learning algorithms possess distinct advantages. Then a combination of multiple machine learning algorithms may improve the robustness of the model's predictive performance.
\section{Acknowledgements}
During manuscript preparation, DeepSeek (V3.0) was employed solely for language polishing and structural suggestions in non-interpretive sections (e.g., methodology descriptions). All AI-processed content underwent expert validation by the authors to ensure adherence to sport science reporting standards.

\end{document}